\documentclass[aps,twocolumn,groupedaddress,showpacs]{revtex4}
\usepackage{graphicx}
\usepackage{amsfonts}
\usepackage{amssymb}
\usepackage{amsmath}

\begin{document}

\title[sinusoidal deformations]
{Suppression of finite-size effects in one-dimensional correlated systems}

\author{A.~Gendiar$^{1,2,3}$, M.~Dani\v{s}ka$^{1,4}$, Y.~Lee$^{1,3}$, and T. Nishino$^3$}
\affiliation{
$^1$ Institute of Physics, Slovak Academy of Sciences, SK-845~11, Bratislava, Slovakia\\
$^2$ Institute of Electrical Engineering, Slovak Academy of Sciences,
SK-841~04, Bratislava, Slovakia\\
$^3$ Department of Physics, Graduate School of Science, Kobe University, Kobe 657-8501, Japan\\
$^4$ Department of Nuclear Physics and Biophysics, Faculty of Mathematics, Physics and Informatics, Comenius University, 
SK-842 48 Bratislava, Slovakia
}

\date{\today}

\begin{abstract}
We investigate the effect of a non-uniform deformation applied to one-dimensional (1D) 
quantum systems, where the local energy scale is proportional to
 $g_j^{~} = [ \sin (j \pi / N) ]^m_{~}$ determined by a positive integer $m$, site index $1 \le j \le N-1$,
and the system size $N$. This deformation introduces a smooth boundary 
to systems with open boundary conditions.
When $m \ge 2$, the leading $1/N$ correction to the ground state energy per
bond  $e_0^{( N )}$ vanishes and one is left with a $1/N^2$ correction, the same as
with periodic boundary conditions.
In particular,  when $m = 2$, the value of $e_0^{( N )}$ 
obtained from the deformed open-boundary system coincides with
the uniform system with periodic boundary conditions.
We confirm the fact numerically for correlated systems, such as the 
extended Hubbard model, in addition to 1D free-Fermion models.
\end{abstract}


\maketitle

\section{Introduction}

The periodic boundary conditions (PBC) are often more convenient than
the open boundary conditions (OBC), when asymptotic properties of
one-dimensional (1D) quantum systems are studied in the thermodynamic limit.
This is partially because boundary energy corrections exist under OBC, 
where eigenstates are not translational invariant. Systems with PBC normally contain 
smaller finite size effects, and this property of PBC is appropriate for accurate determination
of bulk properties by means of the finite size scaling.~\cite{Fisher,Barber,fss}

In numerical studies of lattice models, OBC are often chosen for technical reasons.
In particular, majority of the practical numerical analyses by the density matrix renormalization
group (DMRG) method~\cite{White,Peschel,Schol} are performed under OBC.
Concerning to finite size systems with PBC, the crucial point in DMRG is the ring-shaped
geometry, which reduces the decay rate of the singular values.~\cite{Ver}
Although recent progress in DMRG and the tensor product 
formalisms made it possible to include the PBC in a natural 
manner,~\cite{Ver,Gerd,Rossini1,Rossini2} numerical implementation requires additional 
computational resources compared with conventional DMRG analyses.

A way of suppressing the boundary energy corrections induced by OBC is to introduce
smooth boundary conditions (SBC).~\cite{Vekic,Vekic2} Recently we proposed a variant of
the smooth boundary conditions, where the local energy scale of $N$-site systems is proportional
to the deformation function $[ \sin (j \pi / N) ]^2_{~}$ specified by the site
index $1 \le j \le N-1$.~\cite{ptp1,ptp2} This {\it sine-squared deformation}
(SSD)~\cite{Hikihara} completely suppresses the boundary effects when the ground-state 
energy of a free-Fermions model on the 1D lattice is considered. In this article we generalize the 
deformation function, which is given by $g_j^{~} = [ \sin (j \pi / N) ]^m_{~}$, where $m$ is a positive 
integer.~\cite{Hyperbolic} In the next section, we examine the effect of this {\it sinusoidal 
deformation} by $g_j^{~}$ up to $m = 5$ when it is applied to the 1D free-Fermion model. It is shown 
that the case $m = 2$, the SSD, is the most efficient for the suppression of the boundary 
effect.

Another trial in this article is the application of SSD to correlated systems. As typical 
examples of correlated systems, we choose the spinless-Fermion model with 
nearest-neighbor interaction and the extended Hubbard model; we report the numerical 
result obtained by DMRG in Sec.~III. When the interaction is present, the determination 
of the chemical potential is non trivial. We present a systematic way of solving this 
problem in Sec.~IV. We summarize results in the last section.

\section{Sinusoidal Deformation}

We start from the sinusoidal deformation applied to the free-Fermion model on the 1D lattice.
Consider a tight-binding model represented by the Hamiltonian
\begin{equation}
{\cal H}^{(N)}_{~} = - t \sum_{j=1}^{N-1}  
\left( c^{\dagger}_{j} c^{~}_{j+1} + c^{\dagger}_{j+1} c^{~}_{j} \right)
-\alpha \, t
\left( c^{\dagger}_{N} c^{~}_{1} + c^{\dagger}_{1} c^{~}_{N} \right) \, ,
\label{Hdf}
\end{equation}
where $N$ is the system size, and $t$ the hopping energy. Operators
$c_j^\dagger$ and $c_j^{~}$ represent creation and annihilation  of Fermions, 
respectively.  The parameter $\alpha$ specifies the boundary condition, 
where OBC and PBC correspond to $\alpha = 0$ and $\alpha = 1$, respectively. 
(The choice $\alpha = -1$ is known as the anti-periodic boundary conditions, 
which we do not treat in this article.) For each boundary condition, the one-particle 
energy is expressed as

\begin{equation}
\varepsilon_\ell^{~} =\left\{
\begin{array}{lll}
 -2t \cos{\displaystyle \left(\frac{\pi \ell}{N+1}\right)} & \quad {\rm for\ OBC} & ( \alpha=0 ) \, , \\
& \\
 -2t \cos{\displaystyle \left(\frac{2\pi \ell}{N}\right)} & \quad {\rm for\  PBC} & ( \alpha=1 ) \, , \\
\end{array}
\right.
\end{equation}
where the energy index $\ell$ runs from $1$ to $N$. The ground-state energy $E_0^{(N)}$ 
at half filling is obtained as the sum of $\varepsilon_\ell^{~}$ below the Fermi
energy $\varepsilon_{\rm F}^{~} = 0$. 

Throughout this article we focus on the system size dependence on the
energy per site $E_0^{(N)} / N$ or the energy per bond, which 
is $E_0^{(N)} / N$ under PBC and is $E_0^{(N)} / ( N - 1 )$ under OBC.
After a short algebra, one obtains
\begin{equation}
\frac{E_0^{(N)}}{N} =  
- \frac{2t}{\pi} + \frac{t}{N} \left( 1 - \frac{2}{\pi} \right) 
+ {\cal O} \left( \frac{1}{N^2} \right)
\end{equation}
with OBC, where the leading order of the finite size correction is
proportional to $N^{-1}_{~}$. The correction of the same order also exists for the
energy per bond. With PBC, one finds
\begin{equation}
\frac{E_0^{(N)}}{N} =   
- \frac{2t}{\pi} + \frac{2\pi t}{3N^2} 
+ {\cal O} \left( \frac{1}{N^3} \right) \, ,
\end{equation}
where the leading correction is of the order of $N^{-2}$. The difference between Eq.~(3)
and Eq.~(4) chiefly comes from the presence of the boundary energy which exists only
when OBC are imposed.

The sinusoidal deformation introduces a position dependent energy scale
 $g_j^{~} = \bigl[ \sin (j \pi / N) \bigr]^m_{~}$ to each bond of the system with OBC,
where $m$ is the positive integer.~\cite{ptp1,ptp2} Deforming ${\cal H}^{(N)}_{~}$ in Eq.~(1),
we obtain the corresponding free-Fermionic Hamiltonian
\begin{equation}
{\cal H}^{(N)}_{\rm sine} = -t \sum_{j=1}^{N-1}
\left[ \sin \left(\frac{j \pi }{N}\right) \right]^m_{~}
\left( c^{\dagger}_{j} c^{~}_{j+1} + c^{\dagger}_{j+1} c^{~}_{j} \right) \, .
\label{Hfree}
\end{equation}
We have not obtained analytical solution for 
the one-particle spectrum of ${\cal H}^{(N)}_{\rm sine}$ so far,
except for the zero-energy state. Thus we perform numerical analyses in the
following investigations on the ground state.

Since we are interested in the ground-state energy per site (or per bond), we
introduce the normalization factor
\begin{equation}
B_{~}^{(N)} = \sum_{j=1}^{N-1}
\left[ \sin \left(\frac{j \pi}{N}\right) \right]^m_{~} =
\sum_{j=1}^{N-1} \, g_j^{~} \, ,
\label{nrmfactor}
\end{equation}
which is the sum of the deformation factors over the entire system.
When $m$ is an odd positive integer, we have
\begin{equation}
B_{~}^{(N)}=
\sum\limits_{\ell=0}^{(m-1)/2}\frac{(-1)^\ell}{(2\ell)^{m-1}}
\left(
\begin{array}{c} m\\ \ell \end{array}
\right)
\cot\left[\frac{( m-2\ell )\pi}{2N}\right]
\end{equation}
and when $m$ is an even positive integer, we have
\begin{equation}
B_{~}^{(N)} = 
\frac{N}{2^{m}}\left(\begin{array}{c} m\\m/2 \end{array}\right)\, .
\end{equation}
We represent the ground-state energy of ${\cal H}_{\rm sine}^{(N)}$ at half filling
by the notation $E_0^{(N)}$. It is expected that the normalized energy
\begin{equation}
e_0^{(N)} = \frac{E_0^{(N)}}{B_{~}^{(N)}}
\end{equation}
converges to $- 2 t / \pi$ in the large $N$ limit in analogy to Eqs.~(3) and (4).
We refer $e_0^{(N)}$ in Eq.~(9) to as the energy per bond in the following.
As a convention, we set $B_{~}^{(N)} = N-1$ for the system with OBC,
and $B_{~}^{(N)} = N$ with PBC, where these values just represent
the number of bonds. Using this extended definition of $B_{~}^{(N)}$, we can represent
energy per bond by Eq.~(9) regardless of the boundary condition or the 
presence of deformation.

\begin{figure}
\centerline{\includegraphics[width=0.49\textwidth,clip]{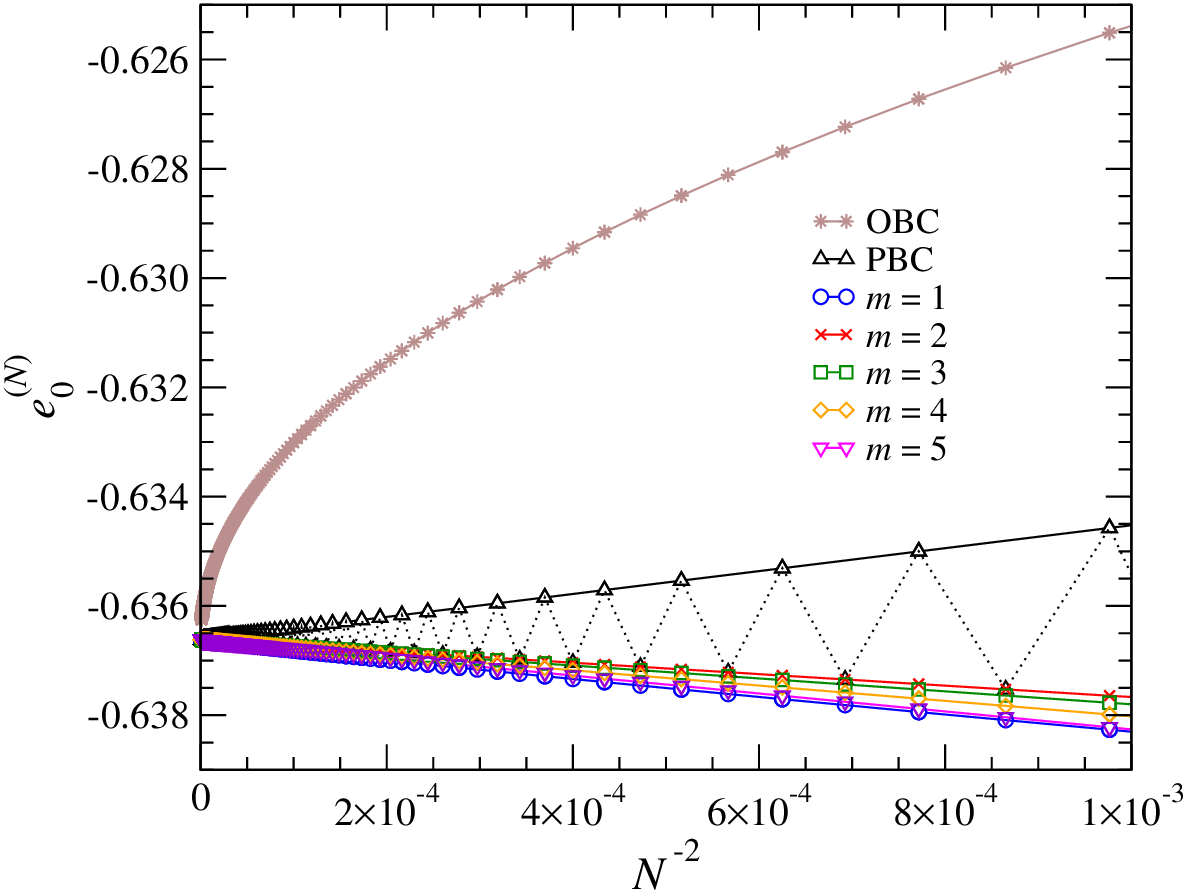}}
\caption{(Color online) Convergence of $e_0^{(N)}$ in Eq.~(9)
with respect to $N$ at half filling. Data with OBC, PBC, and the deformed
cases with $m = 1$ to $5$ are shown.}
\label{fig1}
\end{figure}

We regard $t$ as the unit of the energy in the numerical analyses.
Figure~1 shows the $N$-dependence of  $e_0^{(N)}$ in Eq.~(9) for
the undeformed systems with OBC, PBC, and the deformed systems from $m = 1$ to $5$.
When the PBC are imposed, the convergence of $e_0^{(N)}$ with respect to
$N^{-2}_{~}$ is linear, and there is even-odd oscillation with respect to the 
particle number $N / 2$. Similarly the linear $N^{-2}_{~}$-dependence is observed 
when $m \ge 2$ under the sinusoidal deformation.
In the case $m = 1$, there is additional logarithmic correction as shown later.
It should be noted that when the particle number $N/2$ is odd,
$e_0^{(N)}$ obtained with the sinusoidal deformation
for $m = 2$ coincides with  $e_0^{(N)}$ obtained with PBC.~\cite{AP}
This complete agreement is checked down to the smallest
digit in numerical precision. 
Throughout this paper we use the exact diagonalization in order to reduce
any numerical errors to minimum.

\begin{figure}
\centerline{\includegraphics[width=0.49\textwidth,clip]{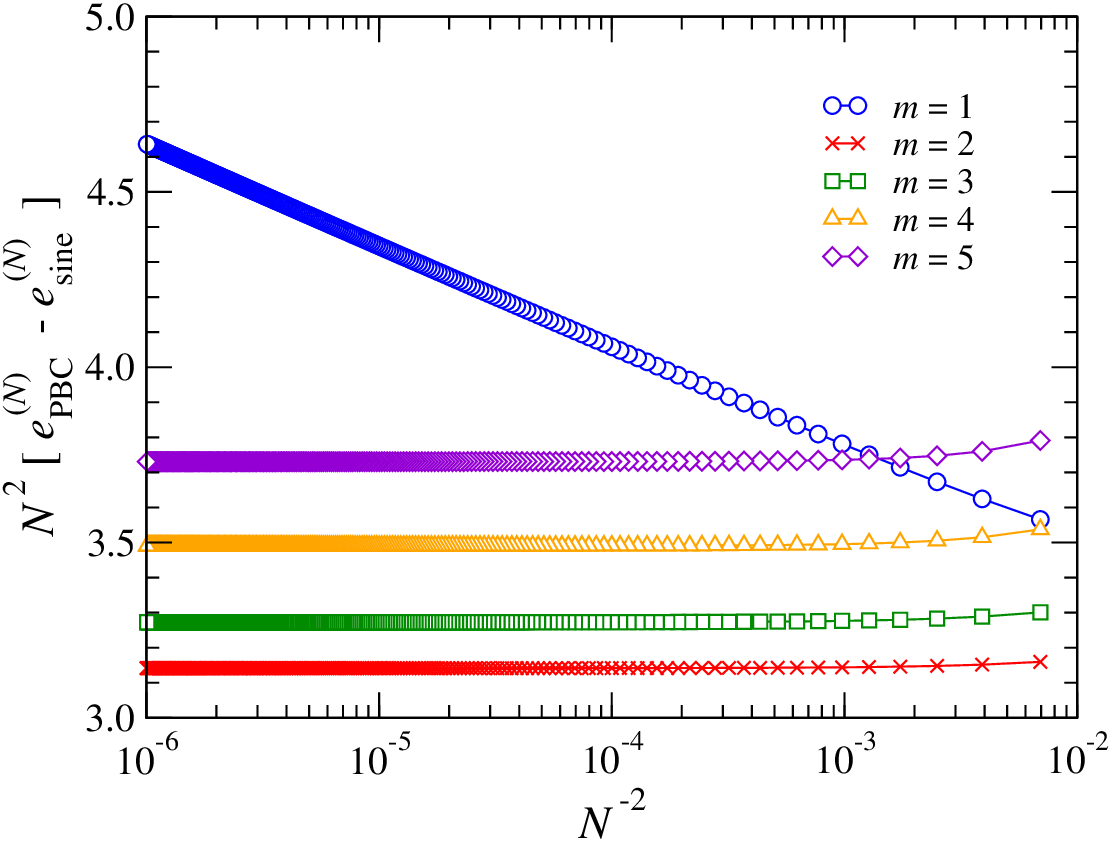}}
\caption{(Color online) Asymptotic behavior of $e_0^{(N)}$ 
for the deformed chains when $1 \le m \le 5$ with respect to  $e_0^{(N)}$ of 
the system with PBC. A logarithmic correction is present when $m = 1$.
}
\label{fig2}
\end{figure}

In order to confirm the $N^{-2}$-dependence of $e_0^{(N)}$ with the sinusoidal deformation
under $m \ge 2$, we plot the difference between $e_0^{(N)}$ obtained with PBC 
(when $N / 2$ is even) and $e_0^{(N)}$ with the sinusoidal deformation. To avoid any confusion, let
$E_{\rm PBC}^{(N)}$ and $E_{\rm sine}^{(N)}$ denote the
ground-state energy obtained with PBC and with the sinusoidal deformation, respectively.
We also use a similar notation for the normalization factors $B_{\rm PBC}^{(N)} = N$ and
 $B_{\rm sine}^{(N)}$ for the normalization factor defined in Eq.~(6). Figure 2 depicts the 
 magnified difference
\begin{equation}
N^2_{~} \left[ e_{\rm PBC}^{(N)} - e_{\rm sine}^{(N)} \right] \equiv
N^2_{~} \left[ 
\frac{E_{\rm PBC}^{(N)}}{B_{\rm PBC}^{(N)}} - 
\frac{E_{\rm sine}^{(N)}}{B_{\rm sine}^{(N)}} \right]
\end{equation}
when $N$ is even. It is shown that the logarithmic correction $(N\log N)^{-1}$ is 
present when $m=1$, and is absent when $m \ge 2$.

\begin{figure}
\centerline{\includegraphics[width=0.49\textwidth,clip]{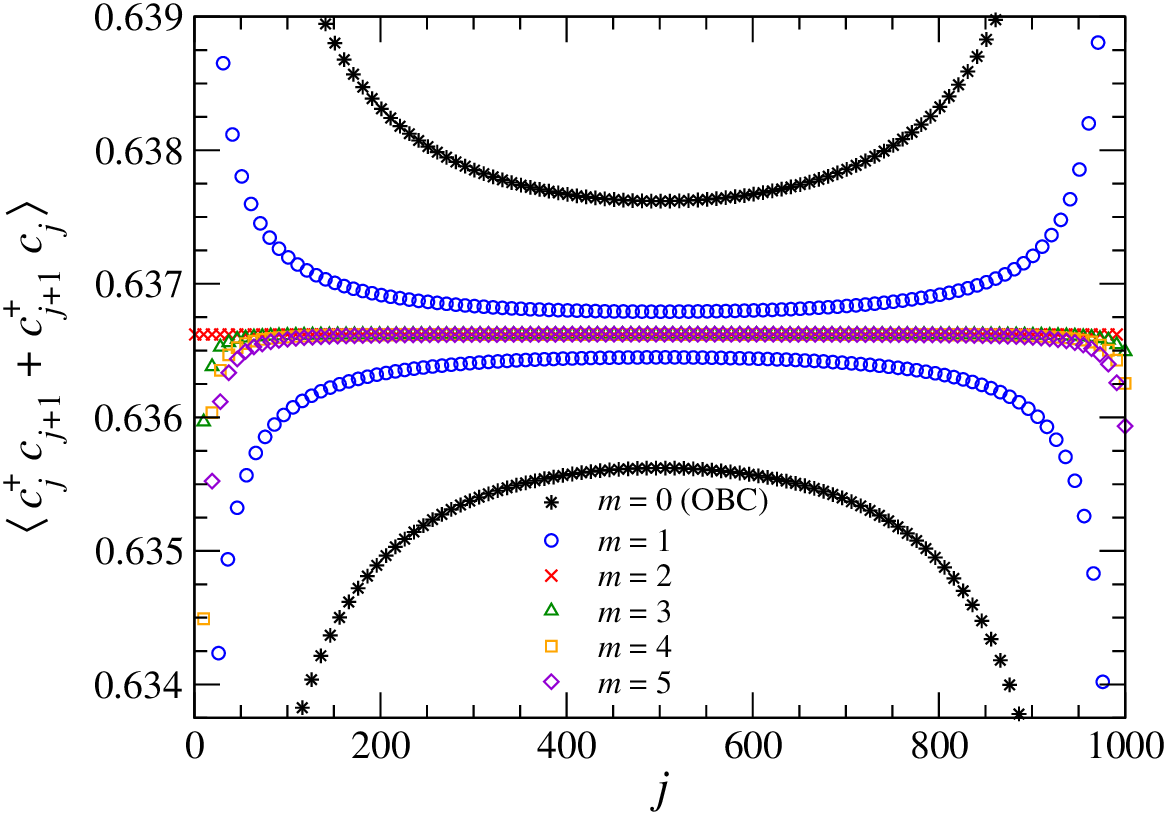}}
\caption{(Color online) Expectation value of the bond correlation function 
 $\langle c_j^\dagger c_{j+1}^{~} + c_{j+1}^\dagger c_j^{~} \rangle$ with respect
to $j$ under sinusoidal deformation. }
\label{fig3}
\end{figure}

Figure 3 shows the spatial distribution of the bond correlation function 
$\langle c_j^\dagger c_{j+1}^{~} + c_{j+1}^\dagger c_j^{~} \rangle$ at half filling 
when $N = 1000$. The Friedel oscillations induced by the boundary are clearly observed when
OBC are imposed (the asterisks), and weaker oscillations are observed with the sinusoidal deformation
when $m = 1$. Only when $m = 2$, there are no oscillations at all; we checked the uniformity
(the translation invariance) of the bond correlation function down to the 16-digits in numerical precision.
When $m \ge 3$, the boundary effects appear again. 
In this case the bond correlation function toward
the system boundary does not oscillate, and decreases in monotonic manner. Such behaviors 
for each $m$ might be related to the suppression of the boundary corrections in $e_0^{(N)}$.

We compare the efficiency of SSD ($m = 2$) with the SBC proposed in Ref.~\cite{Vekic,Vekic2}. 
Figure 4 shows the bond correlation function for both cases at half-filling, where the length of 
boundary area in SBC is chosen as $M = 10$ and $M = 30$ when the system size is $N = 1000$.
Although bulk property is well captured by SBC already for $M = 30$, boundary fluctuations are
still present. On the other hand, the bond correlation function is almost uniform away of the boundary.

\begin{figure}
\centerline{\includegraphics[width=0.49\textwidth,clip]{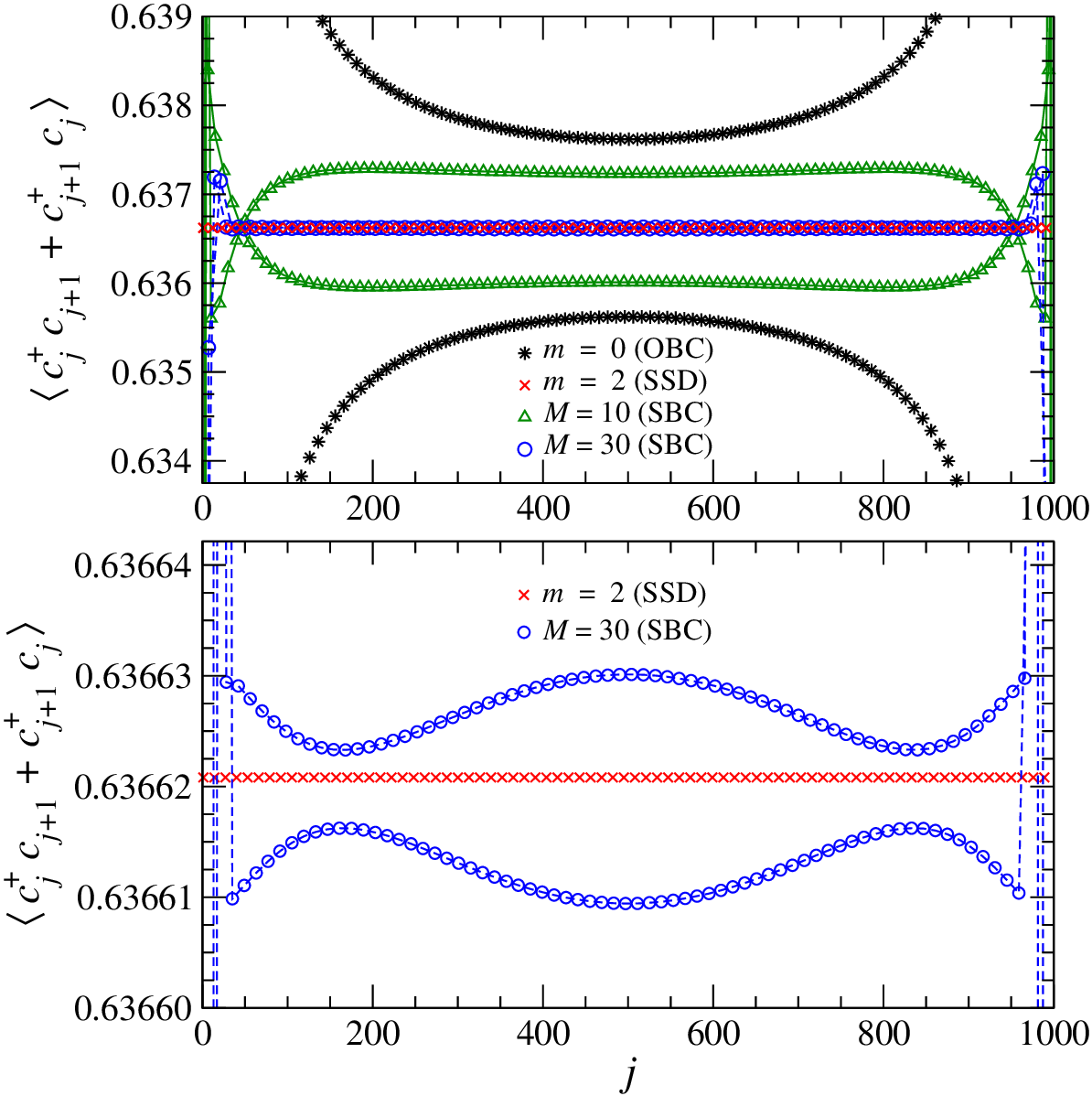}}
\caption{(Color online) Comparison of the expectation value of the bond correlation
function at half filling under SSD ($m = 2$) and SBC in Ref.~\cite{Vekic,Vekic2}. 
The number $M$ in the case of SBC specifies the length of area where the interactions are 
modified near the system boundary.
The bottom graph shows the numerical details when $M=30$ (SBC) with respect to SSD.
}
\label{fig4}
\end{figure}

Now we discuss the way of treating the deformed system away from half filling.
For the undeformed systems with OBC or PBC, it is sufficient to include the chemical
potential term $- \mu \sum_{j=1}^{N} n_j^{~}$ into Eq.~(1), where $n_j^{~} = c_j^\dagger c_j^{~}$
is the number operator. The value of $\mu$ adjusts the Fermi
energy to zero, and is given by
\begin{equation}
\mu( f ) = - 2 t \cos( \pi f ) \, ,
\end{equation}
where $f$ is the filling factor
\begin{equation}
f = \frac{1}{N} \, \sum_{j=1}^N \langle n_j^{~} \rangle \, .
\end{equation}
A natural way of introducing $\mu( f )$ under the sinusoidal deformation is to
write down the Hamiltonian as a sum of the local terms
\begin{equation}
{\cal H}_{\rm sine}^{(N)} = 
\sum_{j=1}^{N-1} \left[ \sin \left(\frac{j \pi}{N}\right) \right]^m_{~} \! h_{j, j+1}^{~} =
\sum_{j=1}^{N-1} g_j^{~} \, h_{j, j+1}^{~} \, \, ,
\end{equation}
where $\mu( f )$ is included to the bond operator
\begin{equation}
h_{j, j+1}^{~} = - t
\left(
c^{\dagger}_{j} c^{~}_{j+1} +  
c^{\dagger}_{j+1} c^{~}_{j}
\right) 
-  \frac{\mu}{2}
\left( n_{j}^{~} + n_{j+1}^{~} \right) \, .
\label{Hsin}
\end{equation}
In order to confirm the validity of these constructions in Eqs.~(11)-(14), 
we carried out numerical calculations for the selected fillings $f = 1/4$ and $f = 1 / 8$.
Figure 5 shows the $N^{-2}_{~}$-dependence
of $e_0^{(N)} = E_0^{(N)} / B_{~}^{(N)}$, where $E_0^{(N)}$ is the
ground-state energy for each filling. We plot the data only when the particle 
number $p \equiv fN$ is even. Analogous to half filling, the bond energy $e_0^{(N)}$ with
PBC coincides with that obtained with SSD ($m = 2$) when $p$ is odd.~\cite{AP}  
The logarithmic corrections are again present when $m = 1$.

\begin{figure}
\centerline{\includegraphics[width=0.49\textwidth,clip]{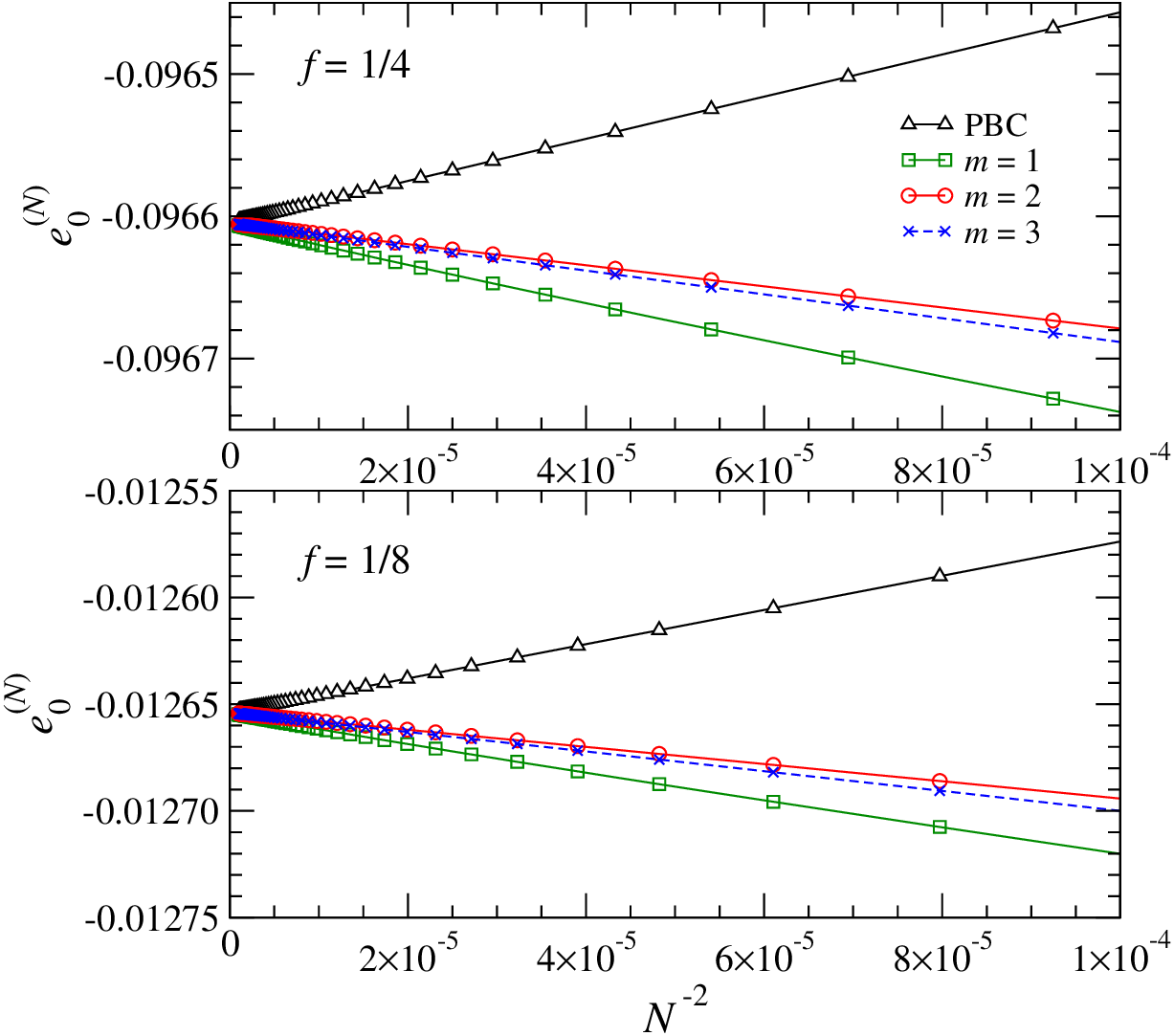}}
\caption{(Color online) 
Convergence of $e_0^{(N)}$ with respect to $N^{-2}_{~}$ at $1/4$-filling and $1/8$-filling.
}
\label{fig5}
\end{figure}

\begin{figure}
\centerline{\includegraphics[width=0.49\textwidth,clip]{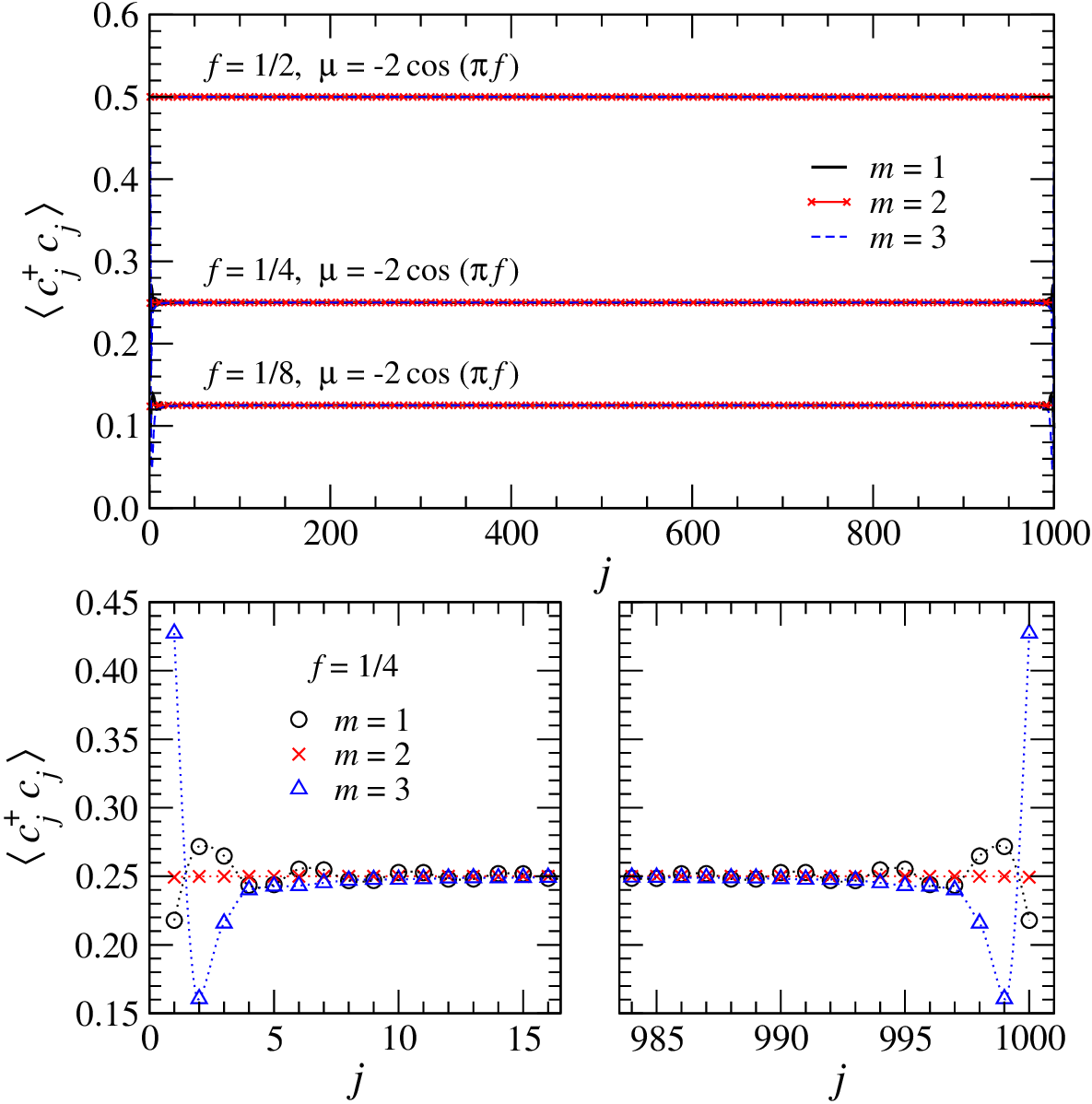}}
\label{fig6}
\caption{
(Color online) Position dependence of the occupation number 
 $\langle n_{j}^{~} \rangle = \langle c^\dagger_j c_j^{~} \rangle$ at $f=1/2$,
$1/4$, and $1/8$ when $m=1$, $2$, and $3$ (upper graph). The left and the
right lower graphs display the occupancy at $f=1/4$ close to the boundaries. }
\end{figure}

The uniformity of the ground state obtained with the sinusoidal deformation is
checked by calculating
the occupancy $\langle n_j^{~} \rangle = \langle c_j^\dagger c_j^{~} \rangle$ at 
$f = 1 / 2$, $1 / 4$, and $1 / 8$. Figure 6 shows $\langle n_j^{~} \rangle$ when 
$m = 1$, $2$, and $3$. At half filling, $f = 1/2$, $\langle n_j^{~} \rangle$ is always equal 
to $1/2$ by the particle-hole symmetry. Even away from the half filling, this uniformity is 
kept when $m = 2$. There are small fluctuations near the system boundary
when $m = 1$ and $m = 3$. In this way, the construction of Hamiltonian in Eqs.~(13) and (14)
away from half-filling is justified, especially for SSD.

\section{Correlated Systems}

It is expected that the SSD reduces boundary effect even when interactions are present 
between particles. In order to check this conjecture, we 
apply the sinusoidal deformation to correlated systems, by means of the Hamiltonian Eq.~(13)
written in the linear combination of bond operators. We study the system size dependence of
the ground-state energy $E_0^{(N)}$ and the uniformity of the system.

\subsection{Spinless Fermions}

As an example, let us consider spinless Fermions on the 1D lattice, whose behavior 
is described by the uniform Hamiltonian
\begin{eqnarray}
{\cal H}^{(N)}_{\rm PBC} = &-& t 
\sum_{j = 1}^N \left(  c_j^\dagger c_{j+1}^{~} + c_{j+1}^\dagger c_j^{~} \right) 
\\
&+& V \sum_{j = 1}^N 
\left( c_j^\dagger c_j^{~} - \frac{1}{2} \right)
\left( c_{j+1}^\dagger c_{j+1}^{~} - \frac{1}{2} \right)
\nonumber
\end{eqnarray}
with PBC, where the system contains the repulsive Coulomb interaction $V > 0$ between 
neighboring sites, in addition to the hopping amplitude $t$. In this section we restrict
ourselves to the half-filled case only, therefore the chemical potential $\mu$ is 
zero.~\cite{XXZ} The construction of the Hamiltonian with OBC and its sinusoidal 
deformation is analogous to what we have done in the previous section. The 
deformed Hamiltonian can be defined by putting the bond operator as 
\begin{eqnarray}
h_{j, j+1}^{~} = &-& t \left(  c_j^\dagger c_{j+1}^{~} + c_{j+1}^\dagger c_j^{~} \right) \\
&+& V 
\left( c_j^\dagger c_j^{~} - \frac{1}{2} \right)
\left( c_{j+1}^\dagger c_{j+1}^{~} - \frac{1}{2} \right) 
\nonumber
\end{eqnarray}
and substituting it to Eq.~(14). We calculate the ground state and the corresponding 
energy $E_0^{(N)}$ at half-filling up to $N = 16$.

\begin{figure}
\centerline{\includegraphics[width=0.49\textwidth,clip]{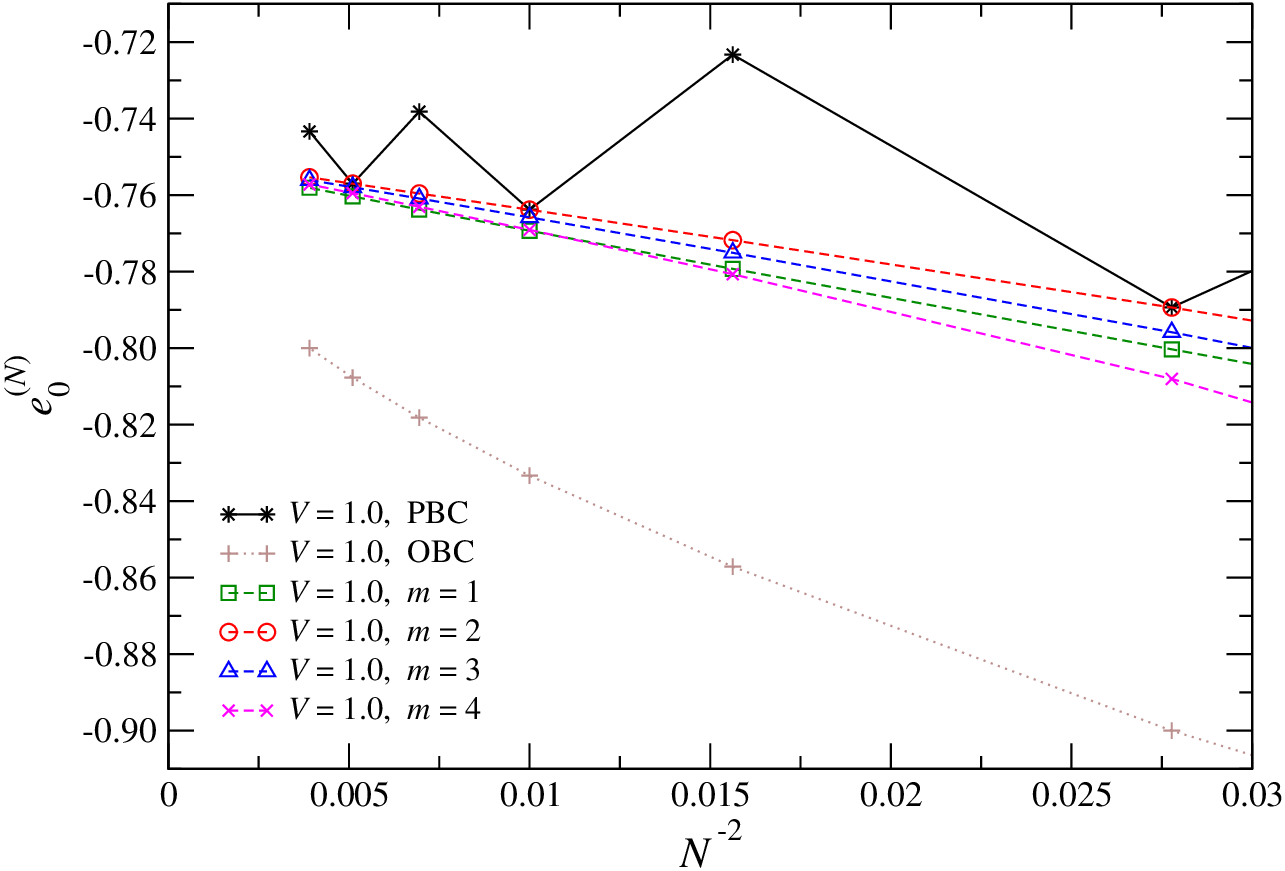}}
\caption{(Color online) 
Energy per bond $e_0^{(N)}$ of the spinless Fermions model at $V = 1$ up to $N = 16$. 
}
\label{fig7}
\end{figure}

Figure~7 shows $e_0^{(N)}$ with respect to $N^{-2}$. 
When the sinusoidal deformation is applied, we observe the same $N^{-2}$ dependence as was
seen for the non-interacting case. In particular when $N/2$ is odd, $e_0^{(N)}$ obtained for
$m=2$ shows a good agreement with the result from PBC.
The coincidence, however, becomes less accurate with increasing $V$, and
there is a deviation around $0.2\%$ in $e_0^{(N)}$ when $V = 5$.
Figure~8 shows the occupation number $\langle n_j^{~} \rangle$ and the bond 
correlation $\langle c_j^{\dagger} c_{j+1}^{~} + c_{j+1}^\dagger c_j^{~} \rangle$ when 
$V = 1$ at half filling.
A clear uniformity is observed when $m = 2$ as shown in the non-interacting cases.

\begin{figure}
\centerline{\includegraphics[width=0.49\textwidth,clip]{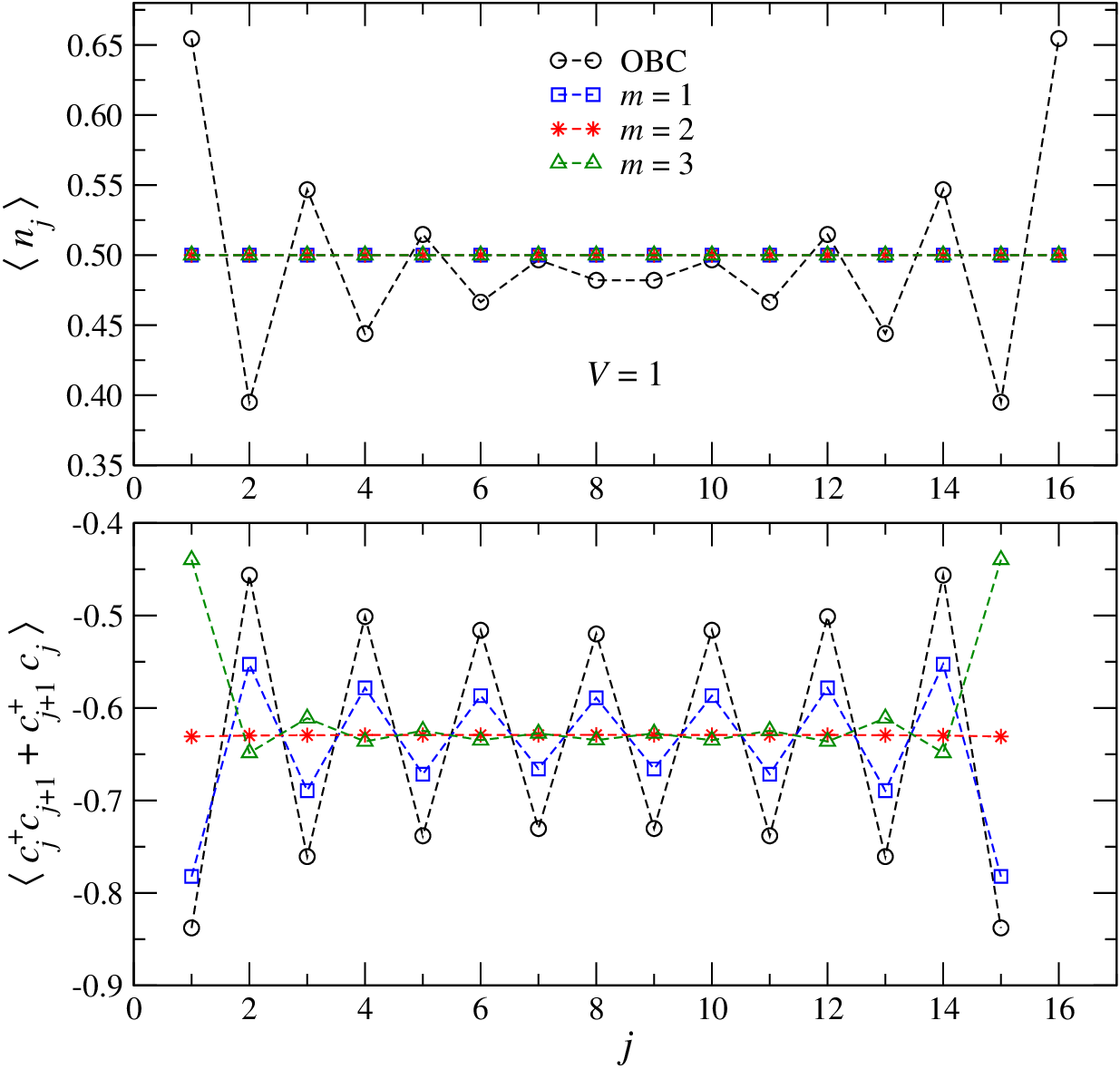}}
\caption{(Color online) 
Occupation number $\langle n_j^{~} \rangle$ and the bond
energy $\langle c_j^{\dagger} c_{j+1}^{~} + c_{j+1}^\dagger c_j^{~} \rangle$ when
$V = 1$ at half filling.}
\label{fig8}
\end{figure}

\subsection{Extended Hubbard Model}

Now we consider the extended Hubbard model, which contains the on-site Coulomb interaction $U$
and the neighboring interaction $V$. In this case there is spin degree of freedom
($\sigma = \uparrow, \downarrow$), therefore creation and annihilation operators, respectively,
are represented as $c_{j \sigma}^{\dagger}$ and $c_{j \sigma}^{~}$. 
The bond operator of the extended Hubbard model is represented as
\begin{eqnarray}
h_{j, j+1}^{~} = &-& t
\sum_{\sigma=\uparrow,\downarrow}^{~}
\left(
c^{\dagger}_{j\sigma} c^{~}_{j+1\sigma} +  
c^{\dagger}_{j+1\sigma} c^{~}_{j\sigma}
\right) 
\nonumber\\ 
&+   & \frac{U}{2}
\left[
\left( n^{\phantom{\dagger}}_{j \uparrow}  -\tfrac{1}{2} \right)
\left( n^{\phantom{\dagger}}_{j \downarrow}-\tfrac{1}{2} \right) \right.
\nonumber\\
&&\left. \quad \quad
+ 
\left( n^{\phantom{\dagger}}_{j+1 \uparrow}  -\tfrac{1}{2} \right)
\left( n^{\phantom{\dagger}}_{j+1 \downarrow}-\tfrac{1}{2} \right)
\right]
\\
& + & V 
\left( n_{j \uparrow}^{~} + n_{j \downarrow}^{~}  - 1 \right)
\left( n_{j+1 \uparrow}^{~} + n_{j+1 \downarrow}^{~} - 1 \right) \, .
\nonumber
\label{Hsin2}
\end{eqnarray}
To avoid the complexity of determining the chemical potential $\mu$, we consider
the half-filled case where $\mu = 0$ is guaranteed by the particle hole symmetry.
Figure 9 shows the $N^{-2}_{~}$-dependence of $e_0^{(N)}$
for various combinations of $U$ and $V$. The coincidence between PBC and SSD
at half filling occurs when the total number of both up- and downs-spin Fermions are odd.

\begin{figure}
\centerline{\includegraphics[width=0.49\textwidth,clip]{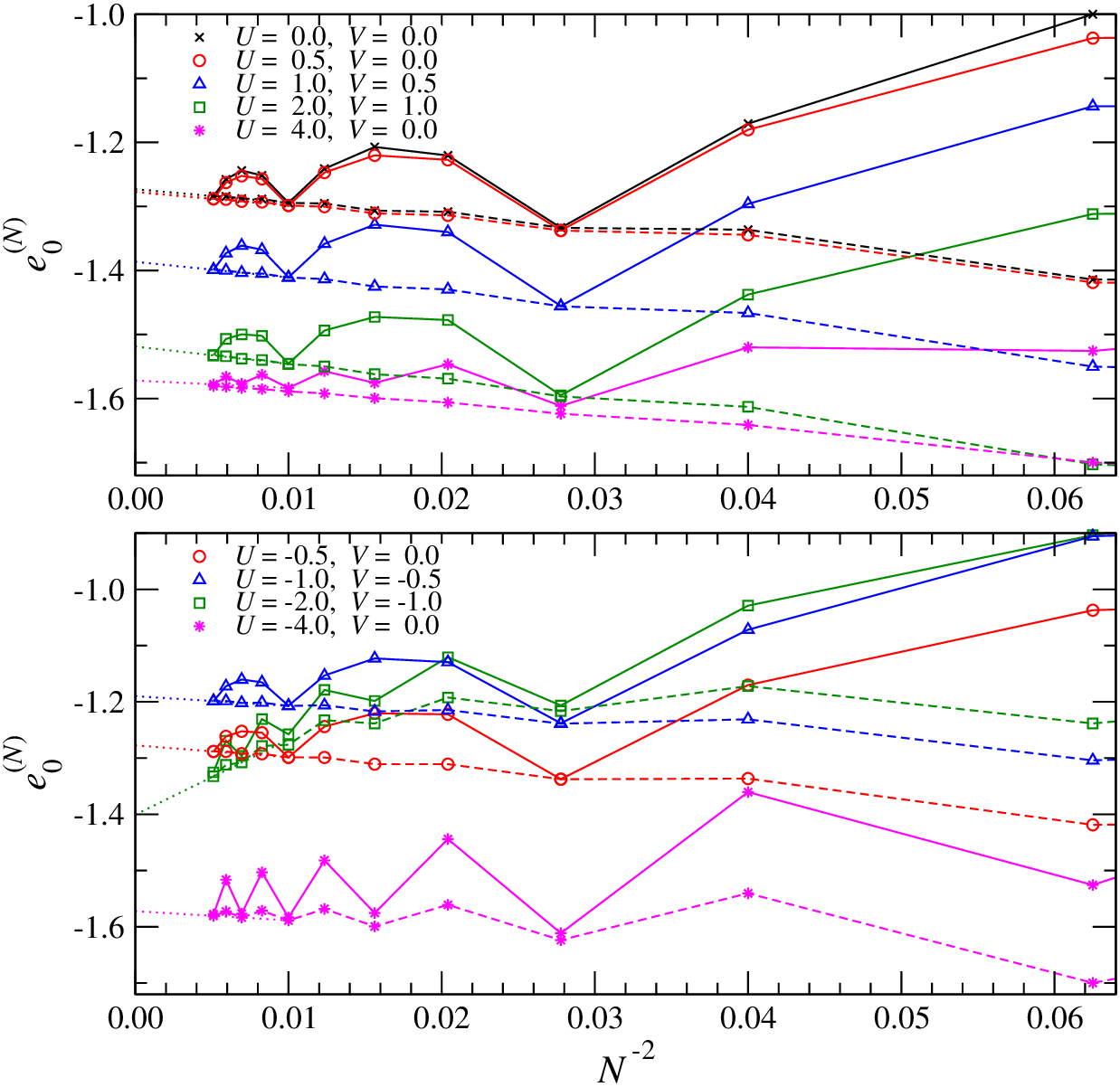}}
\caption{(Color online) System size dependence of $e_0^{(N)}$ of the extended 
Hubbard model at half filling. The full and the dashed lines connect
the energies $e_0^{(N)}$ obtained by PBC and SSD, respectively.}
\label{fig9}
\end{figure}

\begin{figure}
\centerline{\includegraphics[width=0.49\textwidth,clip]{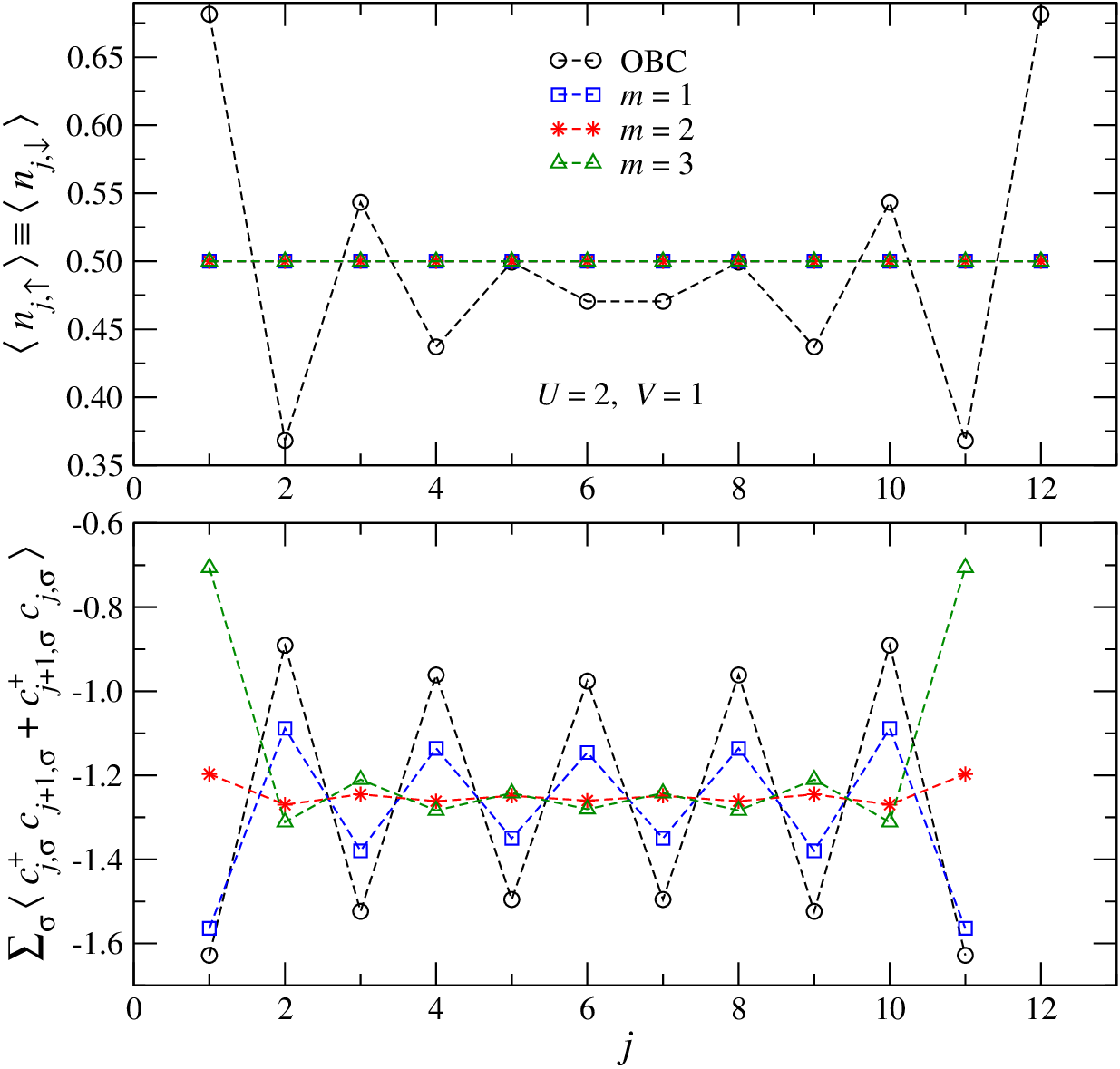}}
\caption{(Color online) 
Occupation $\langle n_{j \uparrow}\rangle = \langle n_{j \downarrow}\rangle$
and the bond energy $\sum_{\sigma}^{~} \langle c^{\dagger}_{j\sigma} c_{j+1}^{~} 
+ c^{\dagger}_{j+1 \sigma} c_{j\sigma}^{~} \rangle$ of the extended Hubbard model 
at half filling.
}
\label{fig10}
\end{figure}

The occupancy $\langle n_j^{~} \rangle = 
\langle n_{j\uparrow}^{~} + n_{j\downarrow}^{~} \rangle$ and
the bond correlation function $\sum_\sigma^{~} \langle c^{\dagger}_{j \sigma} c_{j+1 \sigma}^{~}
+ c^{\dagger}_{j\sigma} c_{j+1\sigma}^{~} \rangle$  at half filling are shown in Fig.~10. 
The occupancy $\langle n_j^{~} \rangle$ with OBC is clearly influenced by the
system boundaries, whereas the sinusoidal deformations ($m\geq1$) lead
to almost constant distribution. The case with $m=2$ (asterisks) realizes the 
minimal position dependence. The bond correlation function is influenced by the 
boundaries in all cases, and the position dependence is the weakest when $m = 2$.

\section{Chemical Potential}

So far we have not discussed the proper value of the chemical potential $\mu$
away from the half filling when the interaction is present.
The chemical potential term $-\tfrac{\mu}{2}(n_{j \uparrow} +
n_{j \downarrow} + n_{j+1 \uparrow} + n_{j+1 \downarrow})$
has to be included to the bond operator in Eq.~\eqref{Hsin2}.
In the Hartree-Fock or the Fermi liquid picture, $\mu$ is adjusted so that the particle number
 $p = \sum_{j=1}^{N}\langle n_{j\uparrow}^{~} + n_{j\downarrow}^{~} \rangle$ 
coincides with the number of negative-energy quasiparticle states.
The number of particles is independently represented by the derivative of $E_0^{(N)}$ 
with respect to $\mu$. Thus, the relation $p = - \partial E_0^{(N)} / \partial \mu$ must
be satisfied for the targeted particle number $p$. Within the sinusoidal deformation,
this relation is slightly modified as
\begin{equation}
p = - \frac{N}{B^{(N)}_{~}} \frac{\partial E_0^{(N)}}{\partial\mu} =
-  \frac{\partial N e_0^{(N)}}{\partial\mu}
\end{equation}
according to the position dependence in the energy scale.

\begin{figure}
\centerline{\includegraphics[width=0.49\textwidth,clip]{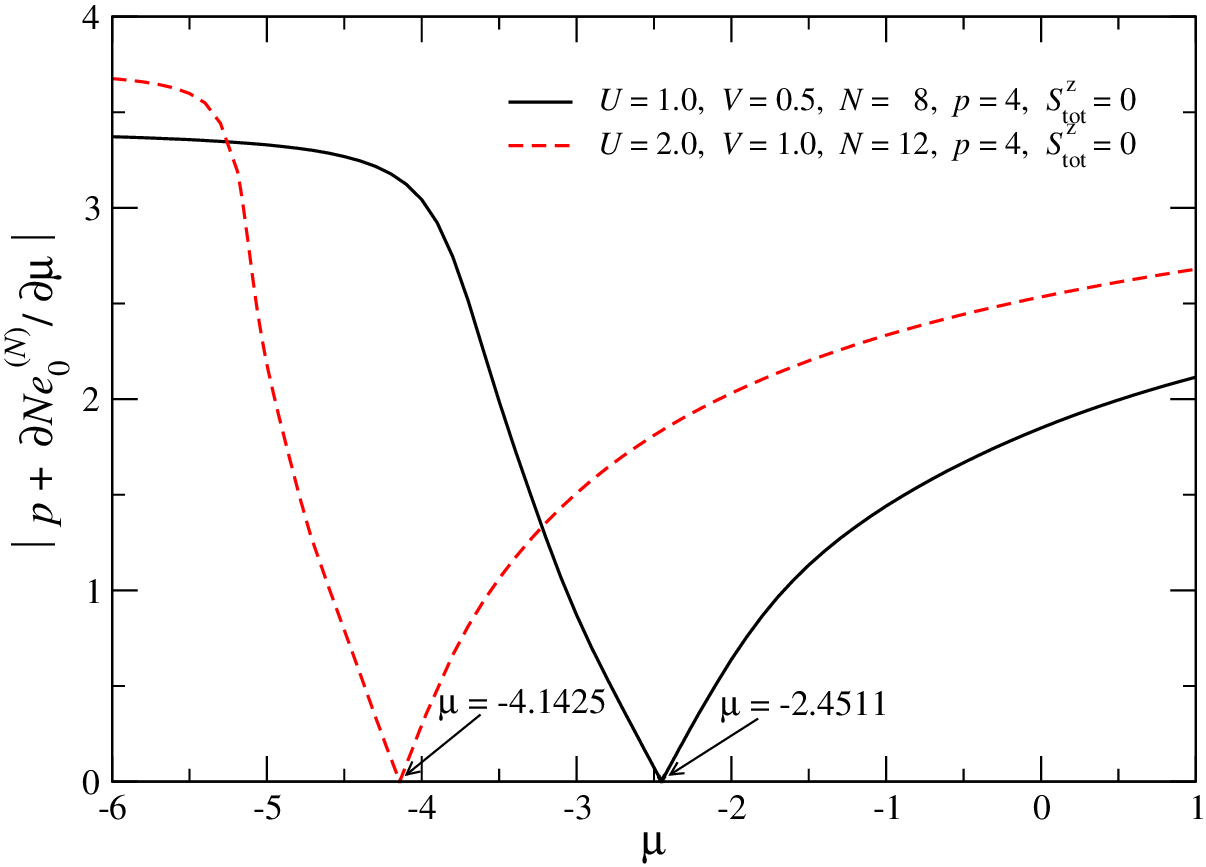}}
\caption{(Color online)
The $\mu$ dependence of $| p + \partial {N e_0^{(N)}} / \partial \mu|$
for the two cases $f \equiv p / 2N = 1 / 4$ and $1 / 6$.}
\label{fig11}
\end{figure}

We plot the absolute value $| p - ( -\partial N e_0^{(N)} / \partial \mu ) |$ with respect to $\mu$ for 
the extended Hubbard model in Fig.~11, under the SSD.
The numerical analysis by exact diagonalization gives $\mu = -4.1425$ for the case
$N = 12$, $p = 4$, $U = 2$, and $V = 1$.
Figure~12 shows the corresponding occupation and the bond correlation functions.
There is a slight position dependence, since we are dealing with relatively small system 
size with a few particles. The position dependence becomes conspicuous
if we choose either an inaccurate value of $\mu$ or we consider the deformations with $m\neq2$.
\begin{figure}
\centerline{\includegraphics[width=0.49\textwidth,clip]{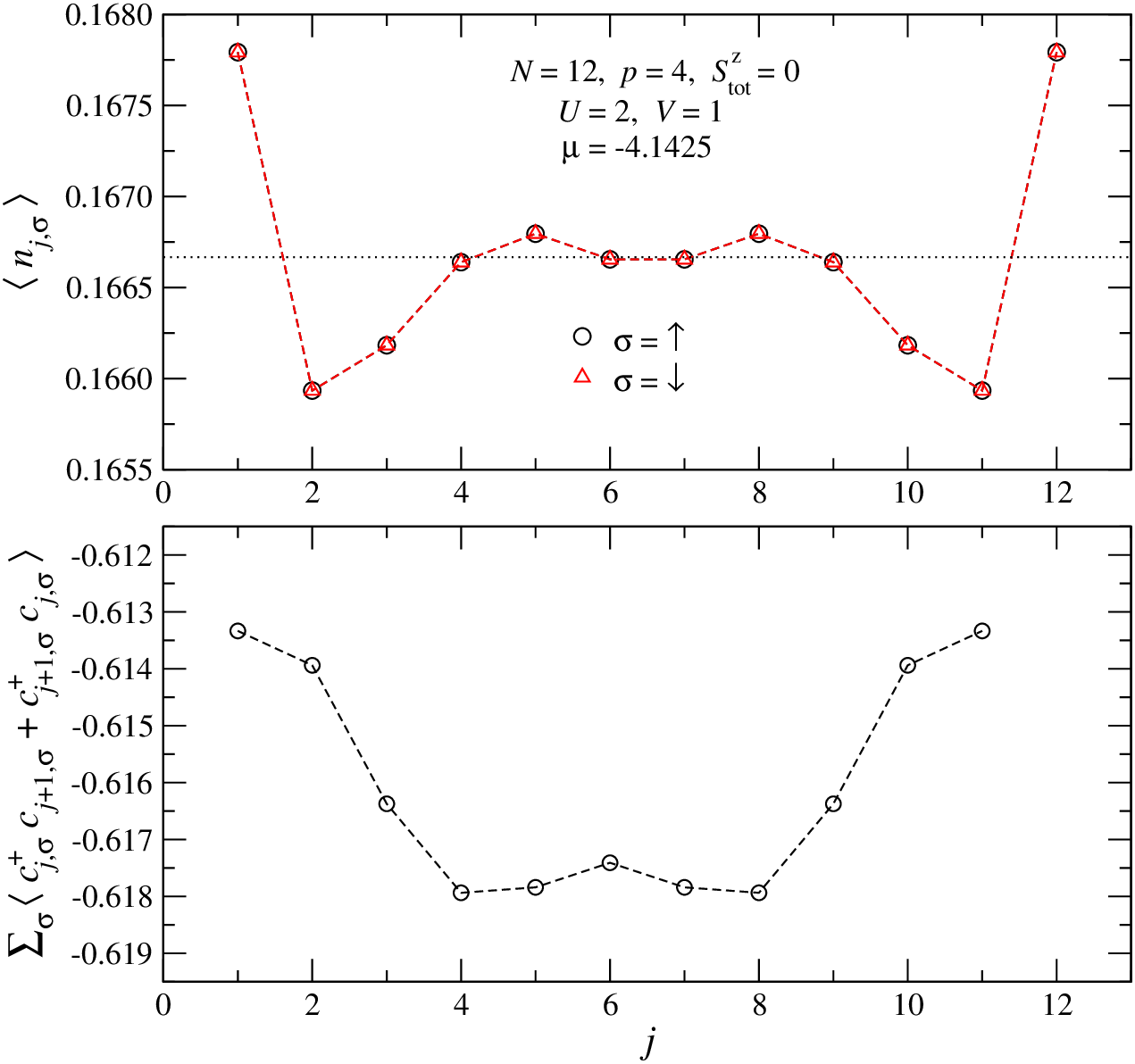}}
\caption{(Color online) 
The upper and lower graphs show the occupancy $\langle n_{i,\sigma}\rangle$ and the 
bond correlation $\sum_{\sigma} \langle c^{\dagger}_{j\sigma} c_{j+1\sigma} + 
c^{\dagger}_{j+1\sigma} c_{j\sigma} \rangle$ at $f=1/6$. }
\label{fig12}
\end{figure}

\section{Conclusion and discussion}

We have shown that the sinusoidal deformation applied to 1D quantum Hamiltonians
improves convergence of the ground state energy per bond (or per site) $e_0^{(N)}$ 
toward the thermodynamic limit, compared to the uniform systems with open
boundary conditions. Such suppression of the boundary effects is confirmed also for interacting
systems, typically for the extended Hubbard model.

We have not determined the `canonical' form of the sinusoidal deformation in case 
where there are long-range interactions. 
It is reported that a small but finite residual boundary effect
appears in spin chains which includes the next-nearest-neighbor interaction.~\cite{Hikihara}
Application of the sinusoidal deformation to higher-dimensional quantum systems 
could be a future problem.

A theoretical puzzle of the sinusoidal deformation remains in coincidence of the ground-state 
energy calculated with PBC and SSD. The agreement is almost perfect, which strongly 
suggests a hidden algebraic relation between these two cases.

\begin{acknowledgments}

We thank T.~Hikihara, A.~Feiguin, and K.~Okunishi for valuable discussions.
This work was supported by ERDF OP R\&D, Project "QUTE - Centre of Excellence
for Quantum Technologies" (ITMS 26240120009), CE QUTE SAV, APVV-51-003505,
VVCE-0058-07, and VEGA-2/0633/09. T.~N. acknowledges the support of 
Grant-in-Aid for Scientific Research (C) No. 22540388.

\end{acknowledgments}


\begin{thebibliography}{99}
\bibitem{AP} If the particle number $N/2$ is even, $e_0^{(N)}$ coincides with 
that obtained for the anti-periodic boundary condition, which is specified by 
$\alpha = -1$ in Eq.~2. The fact was pointed out in Ref.~[16].
\bibitem{XXZ} The Hamiltonian in Eq.~(15) can be mapped to $S = 1/2$ $XXZ$ model 
by means of Jordan-Wigner transformation. The $XXZ$ model under sine-squared 
deformation is studied in Ref.~[16] under the presence of magnetic field, which plays
the role of chemical potential.
\end{thebibliography}
\end{document}